\documentclass{appolb}
\usepackage{graphicx}
\usepackage{cite}
\usepackage{hyperref}
\usepackage{amsmath}
\usepackage{graphicx}
\usepackage{dcolumn}
\usepackage{bm}
\usepackage{enumitem, kantlipsum}
\usepackage{graphicx}

\begin{document}
	
\title{A new and finite family of solutions of hydrodynamics: \\ 
 Part II: Advanced estimate of initial energy densities}
\author{{G. Kasza$^1$ and T. Cs\"org\H{o}$^{1,2}$,}\\[1ex]
	$^1$EKU KRC, H-3200 Gy\"ongy\"os, M\'atrai \'ut 36, Hungary,\\
	$^2$Wigner RCP, H - 1525 Budapest 114, P.O.Box 49, Hungary\\
}	
	
\maketitle
	
\begin{abstract} 
	We derive a new, exact formula for the estimate of the initial
	energy densities from a new family of finite and exact
	solution of relativistic perfect fluid hydrodynamics.
	The new formula depends non-trivially on the speed of sound and on
	the shape or width parameter of the measured (pseudo)rapidity distribution. 
\end{abstract}

\section{Introduction}
This manuscript is the second part of a manuscript series.
The first part deals with the presentation of a recently found, finite and
exact family of solutions of perfect fluid hydrodynamics,
as well as the calculation of the rapidity and the pseudorapidity density
distributions.
The (pseudo)rapidity density distributions can be well described
 by fitting the parameters of the solution to the data,
as described and illustrated in ref.~\cite{CKCJ-WPCF18}. 

This family of solutions can also be used to derive an advanced estimate of the
initial energy densities in high energy  heavy ion and hadron induced
reactions. The precise estimation of these initial energy densities is an
important but difficult problem, and Bjorken's initial energy estimation is one of the top cited results in heavy ion physics.

\section{Bjorken's estimate and its corrections}
The famous formula of Bjorken serves to estimate the initial energy density of a 
longitudinally boost-invariant, cylindrical system. 
It is a simple, phenomenological formula that can be written as:
\begin{equation}\label{eq:Bjorken_estimation}
\varepsilon_0^{Bj}=\frac{\langle m_T \rangle}{R^2 \pi \tau_0}\left.\frac{dN}{dy}\right|_{y=0}.
\end{equation}

This estimate was based on the Hwa-Bjorken solution~\cite{Hwa:1974gn,Bjorken:1982qr}, that corresponds to a flat rapidity density distribution. However, the observed rapidity density is finite, decreases
at large rapidities even at the currently highest CERN LHC energies of 
$\sqrt{s}$ $= $$ 13$ TeV. For
accelerationless, boost-invariant Hwa-Bjorken-flows~\cite{Hwa:1974gn,Bjorken:1982qr},
the initial and final state (indicated by the
$i$ and $f$ indeces) space-time rapidities $\eta_x$ are on the average equal to the rapidity
$y$. Thus for accelerating solutions one has to apply a correction: 
\begin{equation}
\varepsilon_0=\varepsilon_0^{Bj}\frac{dy}{d\eta_x^{f}}\frac{d\eta_x^{f}}{d\eta_x^{i}}.
\end{equation}
The first factor of the correction is the shift of the saddle point corresponding to
the location in $\eta_x$ of the maximum emittivity for particles with a given
rapidity $y$. The second factor describes the increase of the size in  space-time rapidity of a given volume element in accelerating flows. From the 1+1 dimensional, exact and accelerating solution of Cs\"org\H{o}, Nagy and Csan\'ad (CNC) 
these corrections were evaluated and the advanced CNC estimation of the initial
energy density is obtained~\cite{Csorgo:2006ax,Nagy:2007xn} as follows:
\begin{equation}\label{eq:CNC_estimation}
\varepsilon_{0}^{CNC}(\lambda)=\varepsilon_0^{Bj} \left(2\lambda-1\right)\left(\frac{\tau_f}{\tau_0}\right)^{\lambda-1}.
\end{equation}
Here $\lambda \ge 1$ is the acceleration parameter, $\tau_0$ stands for the
thermalization time, corresponding to the initial value of the longitudinal proper time,
and we denoted the final state proper time by
$\tau_f$. 
In the CNC solutions the fluid rapidity is linear in the space-time rapidity
$\eta_x$, given by $\Omega = \lambda \eta_x$, and the Hwa-Bjorken
solution is recovered in the $\lambda \rightarrow 1$ limit.
In this $\lambda = 1$ case,  the velocity field lacks acceleration, 
and the advanced CNC estimation of the initial energy density recovers Bjorken's
formula. 

One of the known problems of this CNC  estimation is that the 1+1 dimensional CNC solution
is valid only for a superhard equation of state,  corresponding to $\kappa = 1 = c_s^2$, 
which is an unrealistic value. To handle this problem, refs.~\cite{Csorgo:2006ax,Nagy:2007xn,Csorgo:2018pxh} proposed a conjecture, that
satisfied 5 criteria:

\begin{enumerate}[noitemsep]
	\item
	It has to reproduce the EoS-independent Bjorken estimate for $\lambda \rightarrow 1$.
\item
	It has to reproduce the exact CNC estimate for any $\lambda$, 
	for $\kappa \rightarrow 1 $.
\item
	It has to follow the known hydro behavior for the energy density $\varepsilon(\tau)$,
	corresponding to exact solutions valid for any (temperature independent) $c_s^2=\frac{1}{\kappa}$. 
	In these solutions an $\varepsilon \propto (\tau_0/\tau)^{ 1/\kappa}$
	behavior is found, assuming that the fluid volume is  proportional to $\tau$.
\item
	It should approximately reproduce the results of numerical hydro calculations; most importantly,
	the additional correction for $\kappa  > 1$ should increase the initial energy density.
\item
	Out of the possible formulas satisfying criteria 1-4, select the simplest one, according to  the
		principle of Occam's razor.
\end{enumerate}

Applying the principle of Occam's razor, the conjectured formula of the initial energy density
that satisfies the above criteria is
\cite{Csorgo:2008pe}:
\begin{equation}\label{eq:CNC_kappadep}
\varepsilon_{0}^{con}(\kappa,\lambda)=\varepsilon_0^{Bj} \left(2\lambda-1\right)\left(\frac{\tau_f}{\tau_0}\right)^{\lambda-1}\left(\frac{\tau_f}{\tau_0}\right)^{\left(\lambda-1\right)\left(1-\frac{1}{\kappa}\right)}.
\end{equation}




According to our expectations, an exact calculation of the initial energy density using an accelerating ($\lambda > 1$) solution with realistic equation of state ($c_s<1$ or $\kappa > 1$), should reproduce the known corrections in the corresponding limits, corresponding to criteria 1, 2 and 3.

Let us test the initial energy density estimates of~\eqref{eq:Bjorken_estimation},~\eqref{eq:CNC_estimation} and~\eqref{eq:CNC_kappadep}, utilizing the
exact results from the CKCJ solution, that has an explicit dependence on the EoS parameter $\kappa$ and the
acceleration parameter $\lambda$. At midrapidity, the average transverse mass per particle is the ratio of the final state energy density $\varepsilon_f$ and particle density $n_f$:
\begin{equation}\label{eq:average_mt}
\langle m_T \rangle = \frac{\left(\frac{dE}{d\eta_x}\right)}{\left(\frac{dN}{d\eta_x}\right)}=\frac{dE}{dN}=\frac{\varepsilon_f}{n_f},
\end{equation}
i.e. the final state energy density can be expressed by the final state particle density as $\varepsilon_f=\langle m_T \rangle n_f$. The rapidity distributions at midrapidity are calculated from the CKCJ solution, using the saddle-point method. In order to express the initial energy density, we used the $\varepsilon(\tau,\eta_x)$ field of the CKCJ solution in the $\eta_x\approx 0$ limit and took the proper time at the final state, $\tau_f$. 
Using the proportionality between $\varepsilon_f$ and $n_f$, we find a new, exact result for the initial energy density:
\begin{equation}\label{eq:CKCJresult}
\varepsilon_0(\kappa,\lambda)=\varepsilon_0^{Bj} \left(2\lambda-1\right)\left(\frac{\tau_f}{\tau_0}\right)^{\lambda\left(1+\frac{1}{\kappa}\right)-1}.
\end{equation}
Rather surprisingly, this formula depends on the equation of state, and this dependence is not vanishing in the $\lambda\rightarrow 1$ limit. Due to that, Bjorken's formula~\eqref{eq:Bjorken_estimation} is not reproduced in the accelerationless case, criteria 1 is violated. The $\kappa \rightarrow 1$ limit does not give back the advanced estimation of the CNC solution, so  criteria 2 is violated too. Compared to Bjorken's estimation, our new, exact result is multiplied by two additional factor that need explanation. 

\begin{figure}[!ht]
	\includegraphics[scale=0.64]{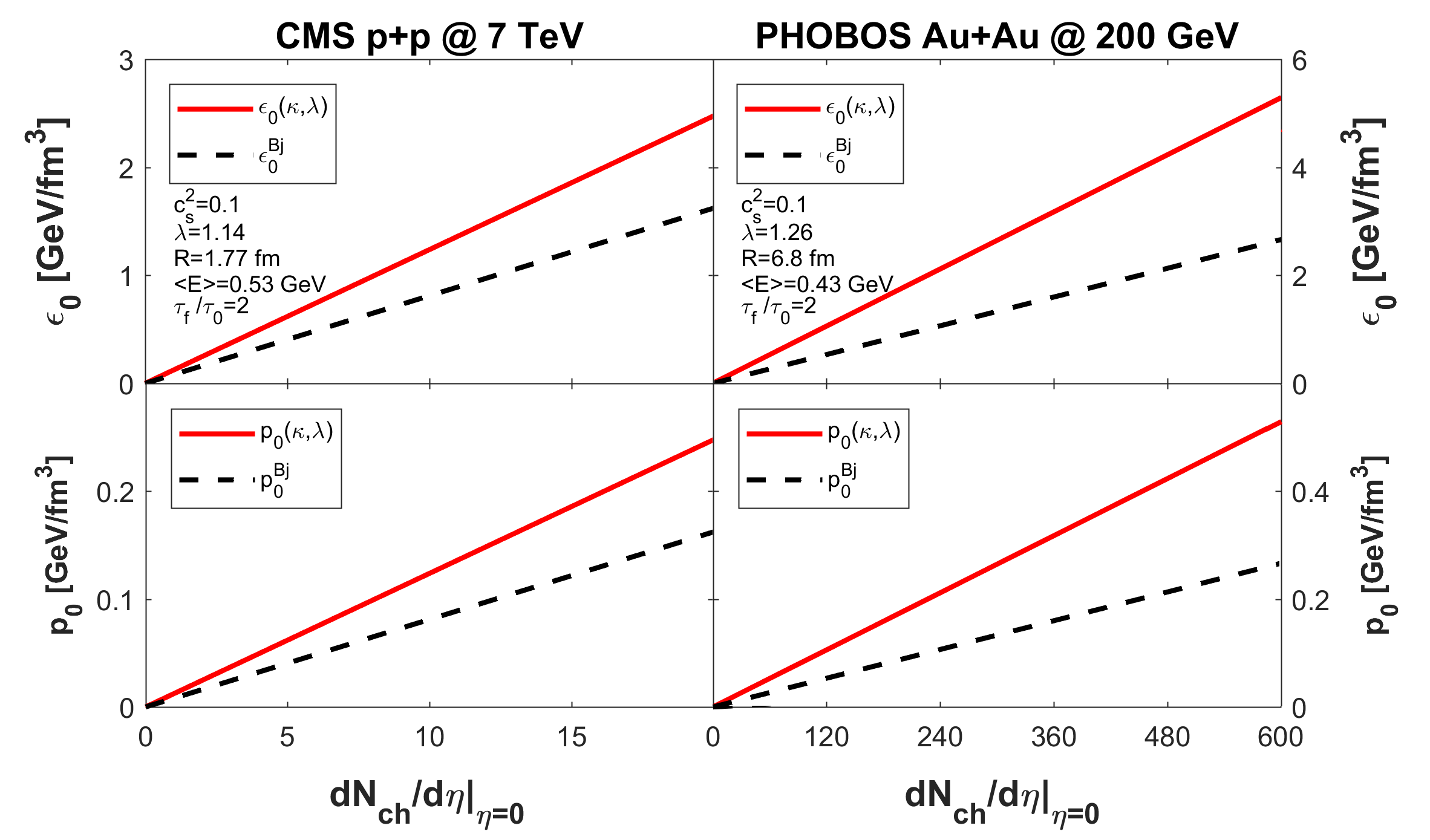}
	\caption{
Initial energy densities (top) and pressures (bottom) from the CKCJ solution are shown with solid lines and compared to Bjorken's estimate, indicated with dashed lines. The parameters of the left panel correspond to fit results of the CKCJ solution  to CMS p+p data at $\sqrt{s} = 7$ TeV, while those of the right panel correspond to similar fits to PHOBOS Au+Au data at $\sqrt{s_{NN}} $$=$$ 200$ GeV. 
	}
	\label{fig:exact_vs_bjorken-a}
\end{figure}
\begin{figure}[!ht]
	\centering
	\includegraphics[scale=0.58]{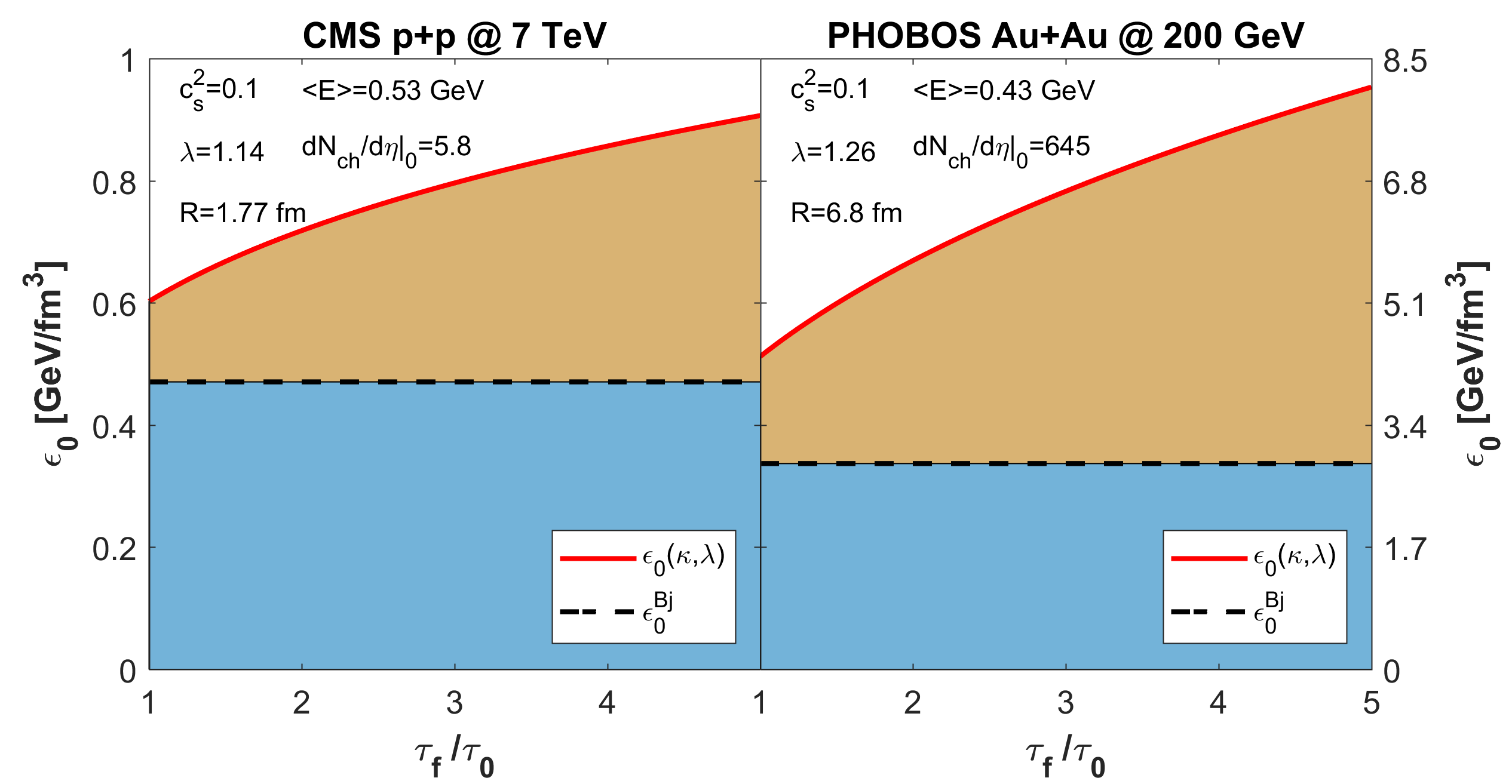}
	\caption{
Initial energy densities (top) and pressures (bottom) from the CKCJ solution are shown with solid lines and compared to Bjorken's estimate, indicated with dashed lines, as a function of the pseudorapidity density of charged particles. The parameters of the left panel correspond to fit results of the CKCJ solution  to CMS p+p data at $\sqrt{s} = 7$ TeV, while those of the right panel correspond to similar fits to PHOBOS Au+Au data at $\sqrt{s_{NN}} $$=$$ 200$ GeV. 
	}
	\label{fig:exact_vs_bjorken-b}
\end{figure}
\begin{figure}[!ht]
	\includegraphics[scale=0.44]{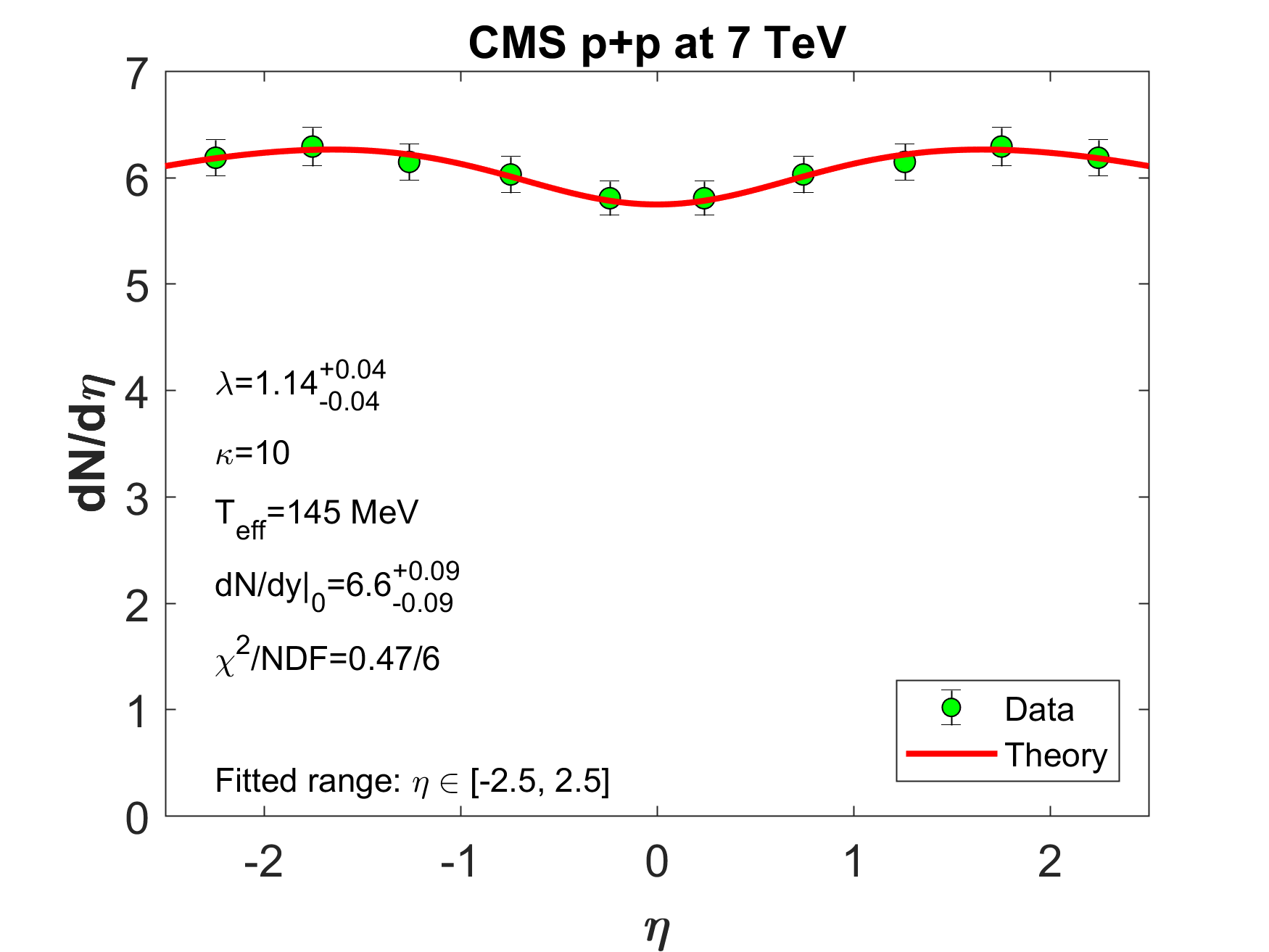}
	\includegraphics[scale=0.44]{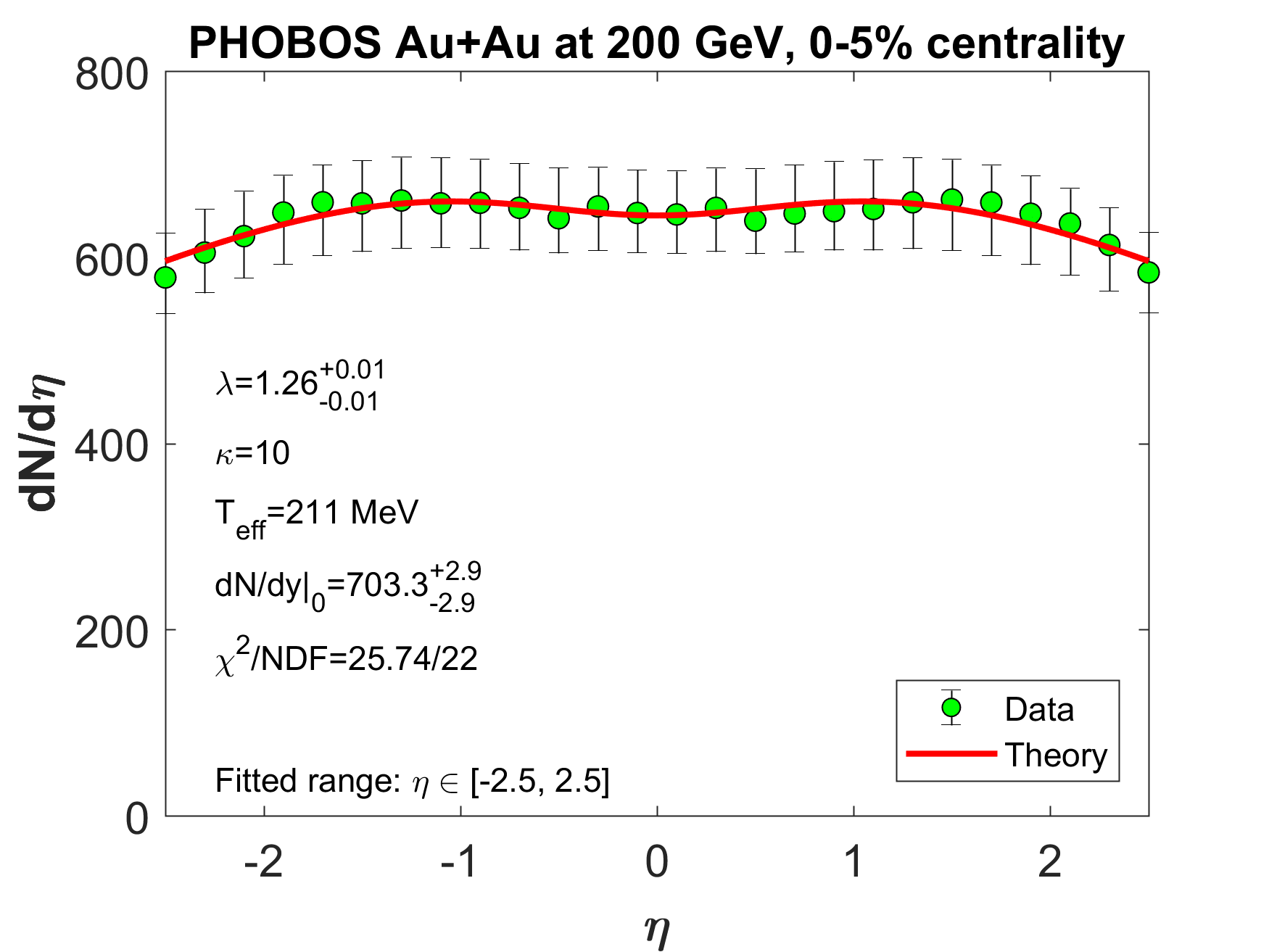}
	\caption{
Left panel shows fits with the CKCJ hydro solution to CMS p+p data at $\sqrt{s} = $ 7 TeV, using $T_{eff} = 145 $ MeV. The right panel is the same, but for PHOBOS Au+Au data at $\sqrt{s_{NN}}$$ =$$ 200$ GeV in the 40-50 \% centrality class, using $T_{eff} = 202$ MeV. The speed of sound is fixed to $c_s^2 = 1/\kappa = 0.1$ in both cases.
}
	\label{fig:ckcj-fits}
\end{figure}

The known $\left(2\lambda-1\right)\left(\tau_f/\tau_0\right)^{\lambda-1}$ factor comes from the shift of the maximum emmitivity and the change of the volume element. In the $\lambda \rightarrow 1$ limit it is vanishing. But the EoS dependent factor is a new correction which violates both the Bjorken and the CNC limit, and it vanishes only for $\kappa\rightarrow \infty$. According to the $p=\varepsilon/\kappa$ equation of state, if $\kappa$ goes to infinity, the speed of sound and the pressure as well becomes zero. Consequently the new, EoS dependent correction factor takes into account the work, which is done by the pressure. Ergo the Bjorken estimation lacks not only the effects of acceleration, but also the initial energy, that is converted to the work of the pressure. Thus Bjorken's estimate is valid only for dust, it neglects the second term of the fundamental thermodynamic relation: $dE=TdS-pdV$.
To visualize the effect of not just the work, but also the acceleration, in Fig.~\ref{fig:exact_vs_bjorken-a} we show the initial energy density $\varepsilon_0$ and the initial pressure $p_0=\varepsilon_0/\kappa$ as functions of the pseudorapidity density of charged particles for our, exact result and for Bjorken's case. In Fig.~\ref{fig:exact_vs_bjorken-b}, we compared the same quantites through their $\tau_f/\tau_0$ dependence. Here we colored the contribution of the Bjorken estimation to blue. We used yellow color for the contribution of the corrections that come from the consideration of the work and the acceleration. For Fig.~\ref{fig:exact_vs_bjorken-a} and Fig.~\ref{fig:exact_vs_bjorken-b} the acceleration parameter is determined by fits of the pseudorapidity density of the CKCJ solution to PHOBOS Au+Au data at $\sqrt{s_{NN}}$$ =$$ 200$ GeV \cite{Alver:2010ck} and CMS p+p data at $\sqrt{s} = $ 7 TeV \cite{Khachatryan:2010us}. These fits are shown on Fig.~\ref{fig:ckcj-fits}.

Our exact result takes into account not only the lack of the rapidity plateaux and the related acceleration of the fluid, but also  the work done  by the pressure during the expansion. The Bjorken estimation clearly underestimates the initial energy density, apparently it lacks both work and acceleration effects. 

The exact result of the CKCJ solution and the conjectured formula of the CNC solution are compared through their ratio:
\begin{equation}\label{eq:exact_conject_ratio}
\frac{\varepsilon_{0}(\kappa,\lambda)}{\varepsilon_{0}^{con}(\kappa,\lambda)}=\left(\frac{\tau_f}{\tau_0}\right)^{\frac{2\lambda-1}{\kappa}-\lambda+1}.
\end{equation}
If we take the $\lambda$ and the $\kappa$ parameter from Fig.~\ref{fig:ckcj-fits}, then according to the fits to CMS p+p data at $\sqrt{s} = $ 7 TeV, the difference between the conjectured formula and the new, exact result  of the initial energy density is only 2\%. The fits to PHOBOS Au+Au data at $\sqrt{s_{NN}}$$ =$$ 200$ GeV predicts more signi\-ficant deviation, it's almost 12\%.  Thus the conjecture of the CNC solution is numerically  surprisingly precise, but pro forma it is inaccurate.

\section{Summary}

A new family of analytic and accelerating, exact and finite solutions of relativistic, perfect fluid hydrodynamics for 1+1 dimensional expanding fireball has been found recently by Cs\"{o}rg\H{o}, Kasza, Csan\'ad and Jiang~\cite{CKCJ-WPCF18,Csorgo:2018pxh}.

This new family of solutions generalizes the 1+1 dimensional Cs\"org\H{o}-Nagy-Csan\'ad solutions for realistic equation of state. With the new solution, the initial energy density has been evaluated. The results were compared to Bjorken's and CNC's initial energy density estimation~\cite{Bjorken:1982qr,Csorgo:2006ax,Csorgo:2008pe}
and to the CNC conjecture~\cite{Csorgo:2008pe}. 
Our new result supersedes all these earlier formulae.

Further generalizations of the CKCJ solution to three dimensionally expanding fireballs, as well as to solutions with a temperature dependent speed of sound are being explored at the time of closing this manuscript.


\begin{thebibliography}{11}

\bibitem{CKCJ-WPCF18}
		T. Cs\"{o}rg\H{o}, G. Kasza, M. Csan\'ad and Z.~F.~Jiang,
		arXiv:1806.06794 [nucl-th].

\bibitem{Bjorken:1982qr} 
J.~D.~Bjorken,
Phys.\ Rev.\ D {\bf 27}, 140 (1983).

\bibitem{Hwa:1974gn} 
R.~C.~Hwa, 
Phys.\ Rev.\ D {\bf 10}, 2260 (1974).

\bibitem{Csorgo:2006ax} 
  T.~Cs\"org\H{o}, M.~I.~Nagy and M.~Csan\'ad,
  Phys.\ Lett.\ B {\bf 663}, 306 (2008).

\bibitem{Nagy:2007xn} 
M.~I.~Nagy, T.~Cs\"org\H{o} and M.~Csan\'ad,
Phys.\ Rev.\ C {\bf 77}, 024908 (2008).

\bibitem{Csorgo:2018pxh} 
T.~Cs\"org\H{o}, G.~Kasza, M.~Csan\'ad and Z.~F.~Jiang,
arXiv:1805.01427 [nucl-th].

\bibitem{Csorgo:2008pe} 
T.~Cs\"org\H{o}, M.~I.~Nagy and M.~Csan\'ad,
J.\ Phys.\ G {\bf 35}, 104128 (2008).



\bibitem{Khachatryan:2010us} 
V.~Khachatryan {\it et al.} [CMS Collab.],
Phys.\ Rev.\ Lett.\  {\bf 105}, 022002 (2010).

\bibitem{Alver:2010ck} 
B.~Alver {\it et al.} [PHOBOS Collab.],
Phys.\ Rev.\ C {\bf 83}, 024913 (2011).

\bibitem{Adare:2006ti} 
A.~Adare {\it et al.} [PHENIX Collab.],
Phys.\ Rev.\ Lett.\  {\bf 98}, 162301 (2007).

\end{thebibliography}
\end{document}